\documentclass[final,5p,times,twocolumn]{elsarticle}

\usepackage{amssymb}

\usepackage{physics}
\usepackage{amsmath,amssymb,bm,bbm,amsfonts}
\usepackage{dsfont}
\usepackage[dvipsnames]{xcolor}
\usepackage{mathrsfs}
\usepackage{latexsym}
\usepackage{amscd}
\usepackage{slashed}
\usepackage{mathtools}
\usepackage[normalem]{ulem}

\def\be{\begin{equation}}
\def\ee{\end{equation}}
\def\bea{\begin{eqnarray}}
\def\eea{\end{eqnarray}}

\journal{Physics Letters B}

\begin{document}


\begin{frontmatter}

\title{Searching for the dark photon at the Future Circular Lepton Collider}

\author[a]{X. G. Wang}
\author[a]{A. W. Thomas}
\affiliation[a]{organization={ARC Centre of Excellence for Dark Matter Particle Physics and CSSM, Department of Physics, \\University of Adelaide},
            city={Adelaide},
            state={South Australia 5005},
            country={Australia}}

\begin{abstract}
In the context of future electroweak precision measurements at the Future Circular Lepton Collider (FCC-ee), we consider recent proposals aimed at finding signatures of physics beyond the Standard Model. In particular, we focus on recent novel suggestions for very precise direct measurements of $\alpha_{\rm e m}(M_Z^2)$. It is shown that at a level of precision of order $10^{-5}$, the effects of a dark photon may be very significant.
\end{abstract}


\end{frontmatter}

\section{Introduction}

A tremendous amount of effort is currently being devoted to the search for new physics beyond the Standard Model (BSM physics). In this context, a recent publication by Riembau~\cite{Riembau:2025ppc} explained the significance of a precise determination of the electromagnetic coupling at the $Z$-pole, $\alpha_{\rm e m}(M_Z^2)$. At present the uncertainty in this value is limited to around $10^{-4}$ by the challenges associated with calculations of the running of $\alpha_{\rm e m}$ arising through hadronic contributions~\cite{Davier:2010nc}. Given that studies of BSM physics based upon electroweak precision measurements at the FCC-ee~\cite{FCC:2018evy} require a precision of order $10^{-5}$ in $\alpha_{\rm e m}(M_Z^2)$~\cite{Riembau:2025ppc}, one needs to find alternative methods.

Some years ago Janot proposed a different method to extract $\alpha_{\rm e m}(M_Z^2)$ at high precision by measuring the forward-backward asymmetry in $e^+ \, e^- \, \rightarrow \mu^+ \mu^-$ for carefully chosen masses at and near ($\sqrt{s_-} = 87.9$ and $\sqrt{s_+} = 94.3$ GeV) the $Z$-pole~\cite{Janot:2015gjr}. This proposal offered a potential precision in $\alpha_{\rm e m}(M_Z^2)$ of order $3 \, \times \, 10^{-5}$. 

A novel, alternative approach, which was only published this year, appears to provide even higher precision. It involves a comparison of the production cross section of electrons versus positrons or muons in $e^+ \, e^- $ annihilation at angles near the forward direction~\cite{Riembau:2025ppc}. Our aim is to reconsider this proposal in the case where the BSM physics involves a dark photon.

Extensions to the gauge sector of the Standard Model (SM) with extra $U(1)$ gauge fields, either through kinetic mixing (dark photon models)~\cite{Fayet:1980ad, Fayet:1980rr, Holdom:1985ag, Okun:1982xi} or anomaly-free $U(1)'$ charges ($Z'$ model)~\cite{Fayet:1990wx, Leike:1998wr, Langacker:2008yv}, have received considerable attention.
Moreover, the dark photon could also be a viable candidate for dark matter~\cite{Caputo:2021eaa, Nguyen:2024kwy}.
While a recent global QCD analysis provided an indirect hint for its existence~\cite{Hunt-Smith:2023sdz}, direct experimental searches have not, so far at least, found evidence of the corresponding dark photon signal~\cite{LHCb:2019vmc, BaBar:2014zli, CMS:2019buh}. Instead, the BaBar~\cite{BaBar:2014zli} and CMS~\cite{CMS:2019buh} collaborations have placed the strongest constraints on the mixing parameter, $\epsilon \le 10^{-3}$. However, these constraints may be significantly relaxed, depending on the detailed structure of the dark sector~\cite{Abdullahi:2023tyk}. Possible new physics phenomena, such as the anomalies in the value of the muon $g-2$, as well as rare kaon and $B$ meson decays which still need further confirmation, are only sensitive to light dark photons with mass below $10\ {\rm GeV}$~\cite{Pospelov:2008zw, Davoudiasl:2012ag, Wang:2023css}. The exclusion limits on the dark photon parameters from electron-nucleon deep-inelastic scattering (DIS) are relatively weak~\cite{Kribs:2020vyk, Thomas:2021lub, Yan:2022npz}.

The electroweak precision observables (EWPO) measured by the Large Electron-Positron Collider (LEP)~\cite{ALEPH:2005ab, ALEPH:2013dgf} have also played an important role in constraining the dark photon parameters~\cite{Hook:2010tw, Curtin:2014cca, Loizos:2023xbj, Harigaya:2023uhg}. The resulting upper bounds on $\epsilon$ are only of order ${\cal O}(10^{-2})$, because of the limited accuracy of the experimental data. New dedicated precision experiments are eagerly anticipated because of their promise in exploring previously unconstrained regions of the dark photon parameter space. 

The high-luminosity, energy frontier Future Circular Lepton Collider 
(FCC-ee)~\cite{FCC:2018evy,TLEPDesignStudyWorkingGroup:2013myl,  Bernardi:2022hny} has been designed to measure a number of electroweak observables with improvements in precision of orders of magnitude~\cite{Proceedings:2015eho}, which will offer great sensitivity to new BSM physics effects~\cite{Ellis:2015sca}. The fermion pair-production processes at future $e^- e^+$ collider with a general $U(1)_X$ gauge boson have been investigated in Ref.~\cite{Das:2021esm}.

Most recently, a new method was proposed~\cite{Riembau:2025ppc} to directly extract $\alpha(M^2_Z)$ at the FCC-ee  by measuring ${\cal R}_{e^-/e+}(\theta)$ and ${\cal R}_{e^-/\mu^-}(\theta)$ at the $Z$ pole, 
where ${\cal R}_{e^-/l^{\pm}}(\theta)$ refers to the ratio between the number of electrons and the number of leptons ($l^{\pm}$) produced at a fixed angle $\theta$. These quantities can be measured very accurately at the FCC-ee with a precision of $10^{-5}$. 

Given such a high level of precision, we expect that this facility should be able to provide stringent constraints on the dark photon parameters. In this work, we investigate the sensitivities of ${\cal R}_{e^-/e+}(\theta)$ and ${\cal R}_{e^-/\mu^-}(\theta)$ to a dark photon. We calculate the $e^- e^+$ cross section with an $s$-channel Breit-Wigner parameterization for the dark photon contributions. We also investigate the impact of the total decay width of the dark photon on the physical observables.

In Sec.~\ref{sec:dark-photon}, we briefly review the dark photon formalism.
We present the $e^- e^+$ cross section with the inclusion of dark photon in Sec.~\ref{sec:cross-section}.
The dark photon corrections to ${\cal R}_{e^-/e+}(\theta)$ and ${\cal R}_{e^-/\mu^-}(\theta)$ are given in Sec.~\ref{sec:corrections}.
A summary of our conclusions is reported in Sec.~\ref{sec:conclusion}.

\section{Dark photon formalism}
\label{sec:dark-photon}

The dark photon is usually introduced as an extra $U(1)$ gauge boson~\cite{Fayet:1980ad, Fayet:1980rr, Holdom:1985ag}, 
interacting with SM particles through kinetic mixing with hypercharge~\cite{Okun:1982xi}
\begin{eqnarray}
\label{eq:L}
{\cal L} & \supset & 
- \frac{1}{4} F'_{\mu\nu} F'^{\mu\nu} + \frac{1}{2} m^2_{A'} A'_{\mu} A'^{\mu} 
+ \frac{\epsilon}{2 \cos\theta_W} F'_{\mu\nu} B^{\mu\nu} 
\, ,
\end{eqnarray}
where $\theta_W$ is the weak mixing angle and $\epsilon$ is the mixing parameter. 
We use $A'$ and $\bar{Z}$ to denote the unmixed versions of the dark photon and the SM neutral weak boson, respectively. 

By performing field redefinitions and diagonalising the mass-squared matrix, the physical $Z$ and dark photon $A_D$ are
\begin{align}
\label{eq:Z-AD}
Z_{\mu} &= \cos\alpha \bar{Z}_{\mu} + \sin\alpha A'_{\mu} \, , \nonumber\\
A_{D\mu} &= - \sin\alpha \bar{Z}_{\mu} + \cos\alpha A'_{\mu}\, .
\end{align}
The $\bar{Z}-A'$ mixing angle $\alpha$ is given by~\cite{Kribs:2020vyk}
\begin{eqnarray}
\tan \alpha &=& \frac{1}{2\epsilon_W} \Big[ 1 - \epsilon^2_W - \rho^2 \nonumber\\
&& - {\rm sign}(1-\rho^2) \sqrt{4\epsilon_W^2 + (1 - \epsilon_W^2 - \rho^2)^2} \Big] \, , 
\end{eqnarray}
where
\begin{align}
\label{eq:epsW-rho}
\epsilon_W &= \frac{\epsilon \tan \theta_W}{\sqrt{1 - \epsilon^2/\cos^2\theta_W}} ,\nonumber\\
\rho &= \frac{m_{A'}/m_{\bar{Z}}}{\sqrt{1 - \epsilon^2/\cos^2\theta_W}} \, .
\end{align}

The couplings of the physical $Z$ and $A_D$ to SM particles (in unit of $e$) are given by \cite{Kribs:2020vyk}
\begin{align}
C_Z^v &= (\cos\alpha - \epsilon_W \sin\alpha) C_{\bar Z}^v + \epsilon_W \sin\alpha \cot \theta_W C_{\gamma}^v ,\nonumber\\
C_Z^a &= (\cos\alpha - \epsilon_W \sin\alpha) C_{\bar Z}^a \, ,
\end{align}
and 
\begin{align}
C_{A_D}^v &= - (\sin\alpha + \epsilon_W \cos\alpha) C_{\bar Z}^v + \epsilon_W \cos\alpha \cot \theta_W C_{\gamma}^v ,\nonumber\\
C_{A_D}^a &= - (\sin\alpha + \epsilon_W \cos\alpha) C_{\bar Z}^a 
\, ,
\end{align}
where $C_{\bar{Z}}^v \sin 2\theta_W = \{g^e_V, g^u_V, g^d_V\}$ and $C_{\bar{Z}}^a \sin 2\theta_W= \{g^e_A, g^u_A, g^d_A\}$ are the Standard Model weak couplings,
\begin{align}
\{ g^e_V, g^u_V, g^d_V\} &= \{ - \frac{1}{2} + 2 \sin^2\theta_W, \frac{1}{2} - \frac{4}{3}\sin^2\theta_W \, ,- \frac{1}{2} + \frac{2}{3}\sin^2\theta_W\} , \nonumber\\
\{ g^e_A, g^u_A, g^d_A\} &= \{ - \frac{1}{2}, \frac{1}{2}, -\frac{1}{2} \} \, ,
\end{align}
and $C_{\gamma} = \{C^e_{\gamma}, C^u_{\gamma}, C^d_{\gamma}\} = \{-1, 2/3, - 1/3 \}$. 

For abbrevity, we will define
\bea
C^v_{Z,f} &=& v_f\, ,\ \ C^a_{Z,f} = a_f\, ,\nonumber\\
C^v_{A_D,f} &=&  \bar{v}_f\, ,\ \ C^a_{A_D,f} = \bar{a}_f\, .
\eea
%


\section{$e^- e^+$ cross section}
\label{sec:cross-section}

In the scattering of $e^-(p_1) e^+(p_2) \to e^-(p_3) e^+(p_4)$, we can neglect the lepton masses and define
 \begin{align}
 s &= (p_1 + p_2) = 4 E^2_{\rm beam}\, ,\nonumber\\
 t &= (p_3 - p_1) = - 2 E^2_{\rm beam} (1 - \cos\theta) = - \frac{s}{2} (1 - \cos\theta)\, ,
 \end{align}
 where $\theta$ is the scattering angle of the outgoing $e^-$ in the centre-of-mass frame.
 The unpolarised differential cross section is
\be
\frac{d\sigma}{d\Omega} = \frac{1}{64\pi^2 s} \cdot \frac{1}{4}\sum_{\rm spin}|{\cal M}|^2 \ ,
\ee
 where the scattering amplitude receives both $s$-channel and $t$-channel contributions with $\gamma$, $Z$ and $A_D$ exchanges.
 Using a Breit-Wigner parametrisation~\cite{Cahn:1986qf} for the $s$-channel $Z$ boson and dark photon propagators, the amplitude can be written as
  \begin{align}
 &  - i {\cal M} \nonumber\\ 
 &= e^2 \bar{u}(p_3)\gamma_{\mu} v(p_4) \frac{1}{s} \bar{v}(p_2) \gamma^{\mu} u(p_1) \nonumber\\
 & + e^2 \bar{u}(p_3)\gamma_{\mu} (v_e - a_e \gamma_5) v(p_4) \frac{1}{s - M^2_Z + i \sqrt{s} \Gamma_Z(s)} \nonumber\\
 & \ \ \ \ \ \times \bar{v}(p_2) \gamma^{\mu} (v_e - a_e \gamma_5)u(p_1) \nonumber\\
 & + e^2 \bar{u}(p_3)\gamma_{\mu} (\bar{v}_e - \bar{a}_e \gamma_5) v(p_4) \frac{1}{s - M^2_{A_D} + i \sqrt{s} \Gamma_{A_D}(s)} \nonumber\\
 & \ \ \ \ \ \times \bar{v}(p_2) \gamma^{\mu} (\bar{v}_e - \bar{a}_e \gamma_5)u(p_1) \nonumber\\
 & - e^2 \bar{u}(p_3)\gamma_{\mu} u(p_1) \frac{1}{t} \bar{v}(p_2) \gamma^{\mu} v(p_4) \nonumber\\
 & - e^2 \bar{u}(p_3)\gamma_{\mu} (v_e - a_e \gamma_5) u(p_1) \frac{1}{t - M^2_Z} \bar{v}(p_2) \gamma^{\mu} (v_e - a_e \gamma_5) v(p_4) \nonumber\\
 & - e^2 \bar{u}(p_3)\gamma_{\mu} (\bar{v}_e - \bar{a}_e \gamma_5) u(p_1) \frac{1}{t - M^2_{A_D}} \bar{v}(p_2) \gamma^{\mu} (\bar{v}_e - \bar{a}_e \gamma_5) v(p_4)\, ,
 \end{align}
where $\Gamma_{Z(A_D)}(s) = \sqrt{s} \Gamma_{Z(A_D)} / M_{Z(A_D)}$.

In the case that the dark photon only decays to the SM final states, its total width is
\begin{align}
\label{eq:Gamma-AD}
\Gamma_{A_D \to \rm SM} &= \sum_f N_C^f \cdot \frac{M_{A_D} \alpha_{\rm em}}{3} \cdot \left\{  \bar{v}_f^2 \left( 1 + \frac{2 m^2_{f}}{M^2_{A_D}} \right) + \right.\nonumber\\
& \left. \bar{a}_f^2 \left( 1 - \frac{4 m^2_{f}}{M^2_{A_D}} \right) \right\} \sqrt{1 - \frac{4 m^2_{f}}{M^2_{A_D}}}\, ,
\end{align}
where $N_C^f = 1$ for leptons and $N_C^f = 3$ for quarks.
If the dark photon also couples to dark matter particles $\chi$ with coupling $g_{\chi}$ which is typically of ${\cal O}(1)$, its total width could be a few orders of magnitude larger. Taking fermionic dark matter as an example, with interaction ${\cal L}_{\chi} = g_{\chi} \bar{\chi} \gamma^{\mu} \chi A'_{\mu}$, the decay width will receive an additional contribution~\cite{Loizos:2023xbj},
\begin{align}
\Gamma_{A_D \to \chi\bar{\chi}} = \frac{M_{A_D} C^2_{A_D,\chi\bar{\chi}}}{12 \pi} \cdot \left( 1 + \frac{2 m^2_{\chi}}{M^2_{A_D}} \right) \sqrt{1 - \frac{4 m^2_{\chi}}{M^2_{A_D}}}\, ,
\end{align}
where the physical coupling is
\be
C_{A_D, \chi\bar{\chi}} = \frac{g_{\chi} \cos\alpha}{\sqrt{1- \epsilon^2/\cos^2\theta_W}}\, .
\ee

 The differential cross section within the SM has been given in Ref.~\cite{Altarelli:1989hv}. It  can be generalised to
\be
\label{eq:dsigma_ee}
\frac{d\sigma_{ee}}{d\cos\theta} = \sum_{i=1}^{21} \frac{d\sigma^i_{ee}}{d\cos\theta}\, ,
\ee
 where the individual terms are given in~\ref{sec:dsigma}.
 
 For $e^- e^+ \to \mu^- \mu^+$ scattering, the cross section only has $s$-channel contributions,
 \be
 \label{eq:dsigma_mumu}
\frac{d\sigma_{\mu\mu}}{d\cos\theta} = \sum_{i=1}^{6} \frac{d\sigma^i_{ee}}{d\cos\theta}\, .
\ee

\section{Dark photon corrections to ${\cal R}_{e^-/e^+}(\theta)$ and ${\cal R}_{e^-/\mu^-}(\theta)$}
\label{sec:corrections}

Following the suggestion of Riembau~\cite{Riembau:2025ppc}, using bins with  $\cos\theta \in [\theta_i, \theta_{i+1}]$, the ratios ${\cal R}_{e^-/\ell^{\pm}}(s, \theta_i)$ can be defined as
\begin{align}
{\cal R}_{e^-/e^+}(s, \theta_i) &= \frac{\int_{\theta_i}^{\theta_{i+1}} \frac{d \sigma_{ee}}{d\cos\theta} (s, \cos\theta) \cdot d \cos\theta}{\int_{\theta_i}^{\theta_{i+1}} \frac{d \sigma_{ee}}{d\cos\theta} (s, \cos(\pi-\theta)) \cdot d \cos\theta}\, , \nonumber\\
{\cal R}_{e^-/\mu^-}(s, \theta_i) &= \frac{\int_{\theta_i}^{\theta_{i+1}} \frac{d \sigma_{ee}}{d\cos\theta} (s, \cos\theta) \cdot d \cos\theta}{\int_{\theta_i}^{\theta_{i+1}} \frac{d \sigma_{\mu\mu}}{d\cos\theta} (s, \cos\theta) \cdot d \cos\theta}\, .
\end{align}
At the $Z$ pole, $\sqrt{s} = M_Z$, the ratio ${\cal R}_{e^-/\ell^{\pm}}(\theta)$ is assumed to be measured in bins of uniform size $\theta_{i+1} - \theta_i = 0.05$, except for the most forward bin $\cos\theta \in [0.95,0.99]$. The latter has the largest cross section and is therefore expected to show the greatest sensitivity.

\begin{figure*}[!htbp]
\begin{center}
\includegraphics[width=0.8\textwidth]{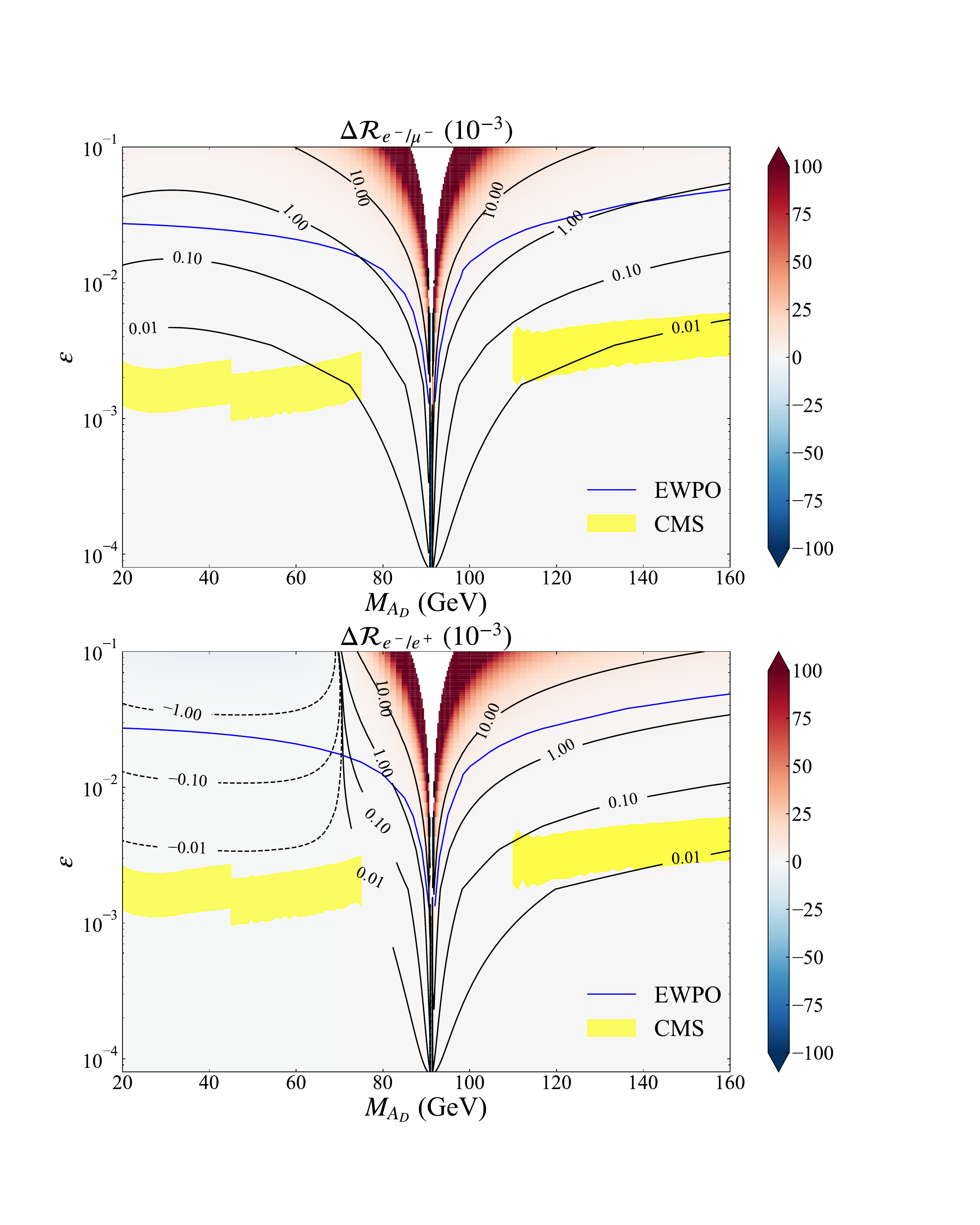}
\vspace*{-0.2cm}
\caption{The dark photon corrections to the SM predictions of ${\cal R}_{e^-/e^+}(\theta_i)$ at the $Z$-pole in the most forward bin $\cos\theta \in [0.95,0.99]$. The gap in the $\epsilon-M_{A_D}$ plane is not accessible due to the ``eigenmass repulsion" associated with the $Z$ boson mass~\cite{Kribs:2020vyk}. The exclusion constraints on $\epsilon$ from electroweak precision observables (EWPO) and the CMS collaboration are taken from Refs.~\cite{Curtin:2014cca} and~\cite{CMS:2019buh}, respectively. The latter assumes only SM coupling and may be significantly relaxed if the dark photon can decay to dark matter~\cite{Felix:2025afw}.
}
\label{fig:Delta_R_p}
\end{center}
\end{figure*}

\begin{figure*}[!htbp]
\begin{center}
\includegraphics[width=0.8\textwidth]{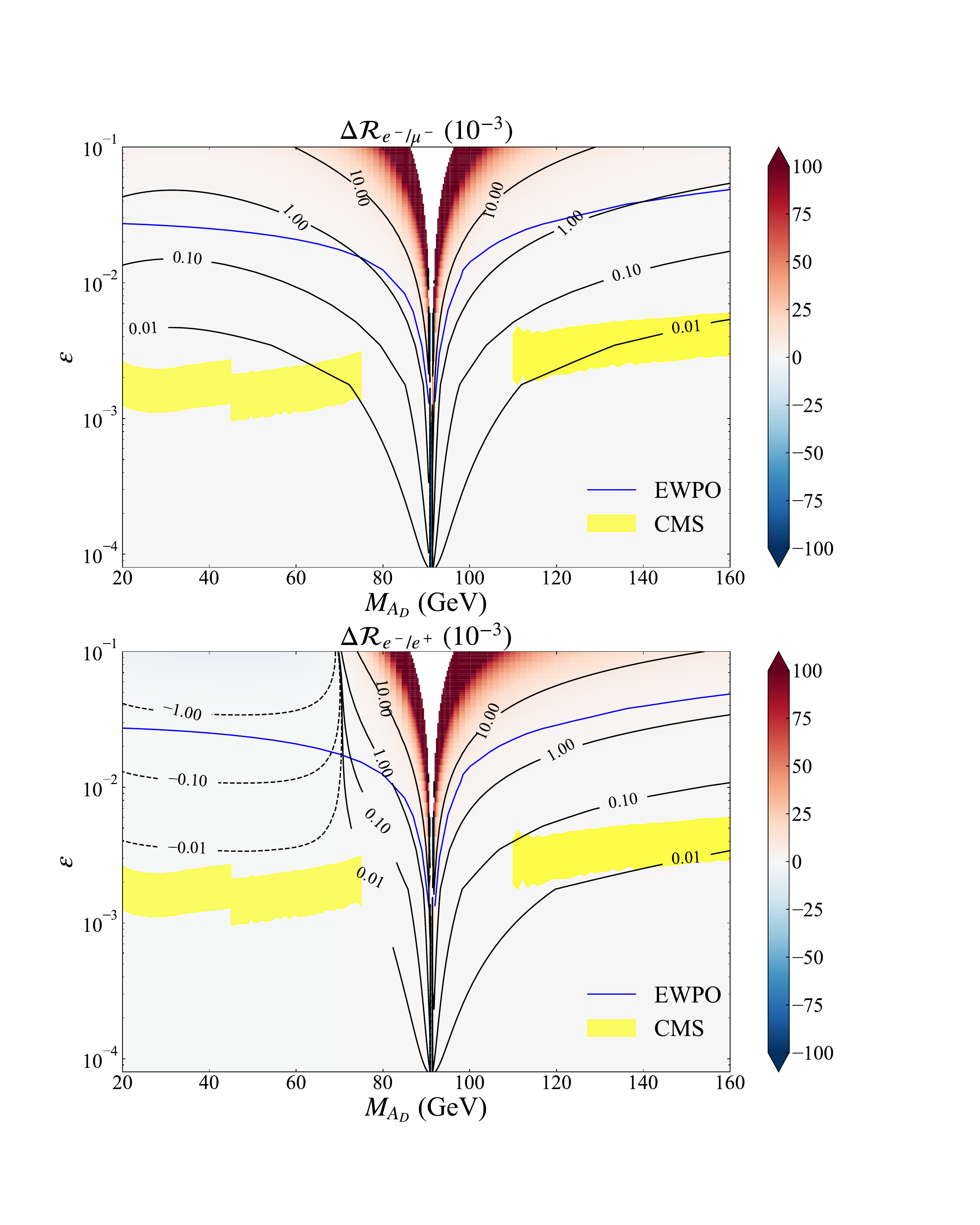}
\vspace*{-0.2cm}
\caption{The dark photon corrections to the SM predictions of ${\cal R}_{e^-/\mu^-}(\theta_i)$ at the $Z$-pole in the most forward bin $\cos\theta \in [0.95,0.99]$.}
\label{fig:Delta_R_mu}
\end{center}
\end{figure*}

The dark photon effect can be characterized by correction factors to the SM predictions,
\bea
{\cal R}_{e^-/e^+}(\theta_i) &=& {\cal R}_{e^-/e^+} (\theta_i)|_{\rm SM} \cdot \Big[ 1 + \Delta {\cal R}_{e^-/e^+}(\theta_i) \Big]\, ,\nonumber\\
{\cal R}_{e^-/\mu^-}(\theta_i) &=& {\cal R}_{e^-/\mu^-}(\theta_i)|_{\rm SM} \cdot \Big[ 1 + \Delta {\cal R}_{e^-/\mu^-}(\theta_i) \Big]\, .
\eea
In our numerical analysis, we use $\alpha^{-1}_{\rm em}(M^2_Z) = 128.952$~\cite{Davier:2010nc}, and $\sin^2\theta_W (M^2_Z)|_{\overline{\rm MS}} = 0.23129$~\cite{ParticleDataGroup:2024cfk}. In Fig.~\ref{fig:Delta_R_p} and Fig.~\ref{fig:Delta_R_mu}, we show the sensitivities of $\Delta {\cal R}_{e^-/e^+}(\theta_i)$ and $\Delta {\cal R}_{e^-/\mu^-}(\theta_i)$ in the most forward bin to the dark photon parameters in the $\epsilon - M_{A_D}$ plane, respectively. These two observables could, in principle, show different sensitivities, as only the former includes a t-channel exchange contribution. In these figures, it has been assumed that the dark photon only decays to SM final states. Within the mass region $M_{A_D} \in [20, 160]\ {\rm GeV}$, the summation in Eq.~(\ref{eq:Gamma-AD}) runs over all SM fermions,  except for the top quark. 

As one sees from Fig.~\ref{fig:Delta_R_p} and Fig.~\ref{fig:Delta_R_mu}, the corrections to the Standard Model predictions may be as large as a few percent when the dark photon parameters approach the ``eigenmass repulsion" region; the region in which one cannot obtain dark photon solutions. 
For a small mixing parameter, even down to $\epsilon \sim 10^{-3}$, the dark photon could lead to relative corrections of order $10^{-5}$,  which are comparable to the anticipated experimental accuracy. 

While the results for both of these processes show a similar sensitivity, it is the latter, which has no t-channel contribution, which looks cleanest. The $e^+ \, e^- \, \rightarrow \mu^+ \, \mu^-$ process is also a golden channel for an accurate measurement of the forward-backward asymmetry, $A^{\mu\mu}_{\rm FB}$. The dominant source of systematic uncertainties comes from the knowledge of the centre-of-mass energy. The total uncertainty is expected to be of order $9\times 10^{-6}$, which is a factor three larger than the statistical error alone~\cite{FCC:2018evy}.

In Figs.~\ref{fig:Delta_R_p} and \ref{fig:Delta_R_mu} we also show the exclusion constraints on $\epsilon$ from electroweak precision observables (EWPO)~\cite{Curtin:2014cca}, which are of ${\cal O}(10^{-2})$. The current upper bounds set by the CMS collaboration are of ${\cal O}(10^{-3})$~\cite{CMS:2019buh}, while the upcoming high-luminosity LHC run (HL-LHC) is expected to place a very strong limit of $\epsilon\le 10^{-6}$ in this mass region~\cite{Curtin:2014cca}.
However, the constraints from these direct experimental searches are model dependent and could be significantly relaxed in light of potential couplings to dark matter particles~\cite{Felix:2025afw}.

We have also checked the impact of a large decay width, arising, for example, from coupling to a dark fermion, by taking $\alpha_D=g^2_{\chi}/4\pi = 0.3$. In the case of $m_{\chi} \ll M_{A_D}$, this leads to $\Gamma_{A_D \to \chi \bar{\chi}} = 4.034\ {\rm GeV}$, compared with $\Gamma_{A_D \to \rm SM} = 0.008\ {\rm GeV}$ for $M_{A_D} = 40\ {\rm GeV}$ and $\epsilon = 0.1$. However, the modifications to both Figs.~\ref{fig:Delta_R_p} and \ref{fig:Delta_R_mu} are negligible.

\section{Conclusion}
\label{sec:conclusion}

We have studied the effect of a massive dark photon on the cross sections for $e^+ \, e^- \, \rightarrow e^+ \, e^-$ and $\mu^+ \, \mu^-$ in the angular regions which are expected to show the greatest sensitivity to $\alpha_{\rm em}(M_Z^2)$. An advantage of the $\mu^+ \, \mu^-$ decay channel is that there is no t-channel contribution. However, both decay modes do exhibit considerable sensitivity to the dark photon contribution. Indeed, it is remarkable that the corrections to the Standard Model predictions can be as large as $10^{-4}$ over the entire dark photon mass region explored, from 20 to 160 GeV, for a mixing parameter $\epsilon$ as small as $10^{-2}$. Very close to the $Z$-boson mass, the corresponding correction can become as large as 1\%. The results presented here suggest that the study of these processes at the FCC-ee could be very important in the search for physics beyond the Standard Model.

\section*{Acknowledgement}

This work was supported by the University of Adelaide and the Australian Research Council through the Centre of Excellence for Dark Matter Particle Physics (CE200100008).


\appendix

\section{Differential cross section}
\label{sec:dsigma}
 \onecolumn
 
 The individual terms in Eq.~(\ref{eq:dsigma_ee}) read
\bea
\frac{d\sigma^1_{ee}}{d\cos\theta} [\gamma(s), \gamma(s)] &=& \frac{\pi\alpha_{\rm em}^2}{2s} \cdot (1 + \cos^2\theta)\, ,\nonumber\\
\frac{d\sigma^2_{ee}}{d\cos\theta} [Z(s), Z(s)] &=& \frac{\pi\alpha_{\rm em}^2}{2s} \cdot \frac{s^2}{(s - M^2_Z)^2 + s \Gamma^2_Z(s)} \cdot \Big[ (v_e^2 + a_e^2)^2 (1+\cos^2\theta) + 8 v_e^2 a_e^2 \cos\theta \Big]\, ,\nonumber\\
\frac{d\sigma^3_{ee}}{d\cos\theta} [\gamma(s), Z(s)] &=& \frac{\pi\alpha_{\rm em}^2}{2s} \cdot \frac{2 s (s-M^2_Z)}{(s - M^2_Z)^2 + s \Gamma^2_Z(s)} \cdot \Big[ v_e^2 (1 + \cos^2\theta) + 2 a^2_e \cos\theta \Big]\, ,\nonumber\\
\frac{d\sigma^{4}_{ee}}{d\cos\theta} [A_D(s), A_D(s)] &=& \frac{\pi\alpha_{\rm em}^2}{2s} \cdot \frac{s^2}{(s - M^2_{A_D})^2 + s \Gamma^2_{A_D}(s)} \cdot \Big[ (\bar{v}_e^2 + \bar{a}_e^2)^2 (1+\cos^2\theta) + 8 \bar{v}_e^2 \bar{a}_e^2 \cos\theta \Big]\, ,\nonumber\\
\frac{d\sigma^{5}_{ee}}{d\cos\theta} [\gamma(s), A_D(s)] &=& \frac{\pi\alpha_{\rm em}^2}{2s} \cdot \frac{2 s (s-M^2_{A_D})}{(s - M^2_{A_D})^2 + s \Gamma^2_{A_D}(s)} \cdot \Big[ \bar{v}_e^2 (1 + \cos^2\theta) + 2 \bar{a}^2_e \cos\theta \Big]\, ,\nonumber\\
\frac{d\sigma^{6}_{ee}}{d\cos\theta} [Z(s), A_D(s)] &=& \frac{\pi\alpha^2_{\rm em}}{2s} \cdot 
\frac{2 s^2 [(s - M^2_Z)(s - M^2_{A_D}) + s \Gamma_Z(s) \Gamma_{A_D}(s)]}{[(s - M^2_Z)^2+ s \Gamma^2_Z(s)] [(s - M^2_{A_D})^2 + s \Gamma^2_{A_D}(s)]} \nonumber\\
&& \cdot \Big[ ( v_e\bar{v}_e + a_e \bar{a}_e )^2 (1 + \cos^2\theta)
+ 2 (v_e \bar{a}_e + a_e \bar{v}_e)^2 \cos\theta \Big]\, ,\nonumber\\
\frac{d\sigma^7_{ee}}{d\cos\theta} [\gamma(t), \gamma(t)] &=& \frac{\pi\alpha_{\rm em}^2}{2s} \cdot \frac{2}{(1 - \cos\theta)^2} \Big[ (1 + \cos\theta)^2 + 4 \Big]\, ,\nonumber\\
\frac{d\sigma^8_{ee}}{d\cos\theta} [Z(t), Z(t)] &=& \frac{\pi\alpha_{\rm em}^2}{2s} \cdot \frac{s^2}{2(t-M^2_Z)^2} \Big\{ (v^2_e + a^2_e)^2 [(1 + \cos\theta)^2 + 4] + 4 v^2_e a^2_e [(1 + \cos\theta)^2 - 4] \Big\}\, ,\nonumber\\
\frac{d\sigma^9_{ee}}{d\cos\theta} [\gamma(t), Z(t)] &=& \frac{\pi\alpha_{\rm em}^2}{2s} \cdot \frac{-2 s}{t - M^2_Z} \Big[ (v_e^2 + a_e^2) \frac{(1 + \cos\theta)^2}{1 - \cos\theta} + 4 (v_e^2 - a_e^2) \frac{1}{1 - \cos\theta} \Big]\, ,\nonumber\\
\frac{d\sigma^{10}_{ee}}{d\cos\theta} [\gamma(s), \gamma(t)] &=& \frac{\pi\alpha_{\rm em}^2}{2s} \cdot \Big[ - \frac{2(1 + \cos\theta)^2}{ 1 - \cos\theta} \Big]\, ,\nonumber\\
\frac{d\sigma^{11}_{ee}}{d\cos\theta} [Z(s), Z(t)] &=& \frac{\pi\alpha_{\rm em}^2}{2s} \cdot \frac{s(s-M^2_Z)}{(s - M^2_Z)^2 + s \Gamma^2_Z(s)} \cdot \frac{s}{t-M^2_Z} \cdot \Big[ (v_e^2 + a_e^2)^2 + 4 v^2_e a^2_e \Big] (1 + \cos\theta)^2\, ,\nonumber\\
\frac{d\sigma^{12}_{ee}}{d\cos\theta} [\gamma(s), Z(t)] &=& \frac{\pi\alpha_{\rm em}^2}{2s} \cdot \frac{s}{t - M^2_Z} \cdot (v^2_e + a^2_e) (1 + \cos\theta)^2\, ,\nonumber\\
\frac{d\sigma^{13}_{ee}}{d\cos\theta} [Z(s), \gamma(t)] &=& \frac{\pi\alpha_{\rm em}^2}{2s} \cdot \frac{-2 s (s-M^2_Z)}{(s - M^2_Z)^2 + s \Gamma^2_Z(s)} \cdot (v^2_e + a^2_e) \cdot \frac{(1 + \cos\theta)^2}{1 - \cos\theta}\, ,\nonumber\\
\frac{d\sigma^{14}_{ee}}{d\cos\theta} [A_D(t), A_D(t)] &=& \frac{\pi\alpha_{\rm em}^2}{2s} \cdot \frac{s^2}{2(t-M^2_{A_D})^2} \Big\{ (\bar{v}^2_e + \bar{a}^2_e)^2 [(1 + \cos\theta)^2 + 4] + 4 \bar{v}^2_e \bar{a}^2_e [(1 + \cos\theta)^2 - 4] \Big\}\, ,\nonumber\\
\frac{d\sigma^{15}_{ee}}{d\cos\theta} [\gamma(t), A_D(t)] &=& \frac{\pi\alpha_{\rm em}^2}{2s} \cdot \frac{- 2 s}{t - M^2_{A_D}} \Big[ (\bar{v}_e^2 + \bar{a}_e^2) \frac{(1 + \cos\theta)^2}{1 - \cos\theta} + 4 (\bar{v}_e^2 - \bar{a}_e^2) \frac{1}{1 - \cos\theta} \Big]\, ,\nonumber\\
\frac{d\sigma^{16}_{ee}}{d\cos\theta} [Z(t), A_D(t)] &=& \frac{\pi\alpha_{\rm em}^2}{2s} \cdot \frac{s^2}{ (t - M^2_Z) (t - M^2_{A_D})} \Big\{ (v_e \bar{v}_e + a_e \bar{a}_e)^2 [(1 + \cos\theta)^2 + 4] + (v_e \bar{a}_e + a_e \bar{v}_e )^2 [ (1+\cos\theta)^2 - 4] \Big\}\, ,\nonumber\\
\frac{d\sigma^{17}_{ee}}{d\cos\theta} [A_D(s), A_D(t)] &=& \frac{\pi\alpha_{\rm em}^2}{2s} \cdot \frac{s(s-M^2_{A_D})}{(s - M^2_{A_D})^2 + s \Gamma^2_{A_D}(s)} \cdot \frac{s}{t-M^2_{A_D}} \cdot \Big[ (\bar{v}_e^2 + \bar{a}_e^2)^2 + 4 \bar{v}^2_e \bar{a}^2_e \Big] (1 + \cos\theta)^2\, ,\nonumber\\
\frac{d\sigma^{18}_{ee}}{d\cos\theta} [\gamma(s), A_D(t)] &=& \frac{\pi\alpha_{\rm em}^2}{2s} \cdot \frac{s}{t - M^2_{A_D}} \cdot (\bar{v}^2_e + \bar{a}^2_e) (1 + \cos\theta)^2\, ,\nonumber\\
\frac{d\sigma^{19}_{ee}}{d\cos\theta} [Z(s), A_D(t)] &=& \frac{\pi\alpha_{\rm em}^2}{2s} \cdot \frac{s(s-M^2_Z)}{(s - M^2_Z)^2 + s \Gamma^2_Z(s)} \cdot \frac{s}{t-M^2_{A_D}} \cdot \Big[ (v_e^2 + a_e^2) (\bar{v}_e^2 + \bar{a}_e^2) + 4 v_e \bar{v}_e a_e \bar{a}_e \Big] (1 + \cos\theta)^2\, ,\nonumber\\
\frac{d\sigma^{20}_{ee}}{d\cos\theta} [A_D(s), \gamma(t)] &=& \frac{\pi\alpha_{\rm em}^2}{2s} \cdot \frac{-2 s (s-M^2_{A_D})}{(s - M^2_{A_D})^2 + s \Gamma^2_{A_D}(s)} \cdot (\bar{v}^2_e + \bar{a}^2_e) \cdot \frac{(1 + \cos\theta)^2}{1 - \cos\theta}\, ,\nonumber\\
\frac{d\sigma^{21}_{ee}}{d\cos\theta} [A_D(s), Z(t)] &=& \frac{\pi\alpha_{\rm em}^2}{2s} \cdot \frac{s(s-M^2_{A_D})}{(s - M^2_{A_D})^2 + s \Gamma^2_{A_D}(s)} \cdot \frac{s}{t-M^2_Z} \cdot \Big[ (v_e^2 + a_e^2) (\bar{v}_e^2 + \bar{a}_e^2) + 4 v_e \bar{v}_e a_e \bar{a}_e \Big] (1 + \cos\theta)^2\, .
\eea
%


\twocolumn


\end{document}